\documentclass[11pt,letterpaper]{article}
\usepackage{jheppub}
\usepackage{color}
\usepackage{graphicx}
\usepackage{wrapfig}
\usepackage{epstopdf}

\usepackage{verbatim}
\usepackage{subfig}
\usepackage{url}
\usepackage{bbold}
\usepackage{xspace}
 \usepackage{setspace}

\usepackage{floatrow}
\newfloatcommand{capbtabbox}{table}[][\FBwidth]

\usepackage{slashed}
\usepackage{verbatim}

\usepackage{amsmath,slashed, amssymb}
\usepackage{fancyhdr}
\usepackage{hyperref}
\usepackage[titletoc]{appendix}
\usepackage{axodraw4j}
\usepackage{pstricks}
\numberwithin{equation}{section} 

\def\beq{\begin{eqnarray}}
\def\eeq{\end{eqnarray}}
\def\bea{\begin{eqnarray*}}
\def\eea{\end{eqnarray*}}

\def\centeron#1#2{{\setbox0=\hbox{#1}\setbox1=\hbox{#2}\ifdim
\wd1>\wd0\kern.5\wd1\kern-.5\wd0\fi
\copy0\kern-.5\wd0\kern-.5\wd1\copy1\ifdim\wd0>\wd1
\kern.5\wd0\kern-.5\wd1\fi}}
\def\ltap{\;\centeron{\raise.35ex\hbox{$<$}}{\lower.65ex\hbox{$\sim$}}\;}
\def\gtap{\;\centeron{\raise.35ex\hbox{$>$}}{\lower.65ex\hbox{$\sim$}}\;}

\newcommand{\newc}{\newcommand}
\newc{\qbar}{{\overline q}}
\newc{\Kahler}{Kahler }
\newc{\sz}{{SUZ$_2$}}
\newc{\zt}{{$\mathbb{Z}_2$}}
\newc{\deltaGS}{\delta_{\rm GS}}
\begin{document}
\title{Strong CP and \sz}

\author[a]{Abdelhamid Albaid}
\author[b]{Michael Dine}
\author[c,b]{Patrick Draper}

\affiliation[a]{University of Hail, Hail, Saudi Arabia}
\affiliation[b]{Santa Cruz Institute for Particle Physics and Department of Physics, University of California, Santa Cruz, CA 93106, USA}
\affiliation[c]{Department of Physics, University of California, Santa Barbara, CA 93106, USA}

\emailAdd{albaid1979@gmail.com}
\emailAdd{mdine@ucsc.edu}
\emailAdd{pidraper@physics.ucsb.edu}

\date{\today}

\abstract{Solutions to the strong CP problem typically introduce new scales associated with the spontaneous breaking of symmetries. Absent any anthropic argument for small $\bar\theta$, these scales require stabilization against ultraviolet corrections. Supersymmetry offers a tempting stabilization mechanism, since it can solve the ``big" electroweak hierarchy problem at the same time. One family of solutions to strong CP, including generalized parity models, heavy axion models, and heavy $\eta^\prime$ models, introduces \zt~copies of (part of) the Standard Model and an associated scale of \zt-breaking. We review why, without additional structure such as supersymmetry, the \zt-breaking scale is unacceptably tuned.  We then study ``\sz" models, supersymmetric theories with \zt~copies of the MSSM. We find that the addition of SUSY typically destroys the \zt~protection of $\bar\theta=0$, even at tree level, once SUSY and \zt~are broken. In theories like supersymmetric completions of the twin Higgs, where \zt~addresses the little hierarchy problem but not strong CP, two axions can be used to relax $\bar\theta$.

}

%\keywords{}

\arxivnumber{1510.03392}

\preprint{}

\maketitle
\section{Introduction}

The upper bound on the neutron electric dipole moment, $|d_n|<2.9\times 10^{26}~e~{\rm cm}$~\cite{Baker:2006ts}, bounds the QCD vacuum angle by
\begin{align}
\bar\theta\equiv (\theta-\arg\det m_q) \lesssim 10^{-10}\;.
\end{align}
More than any other small number problem in high energy physics, the strong CP problem demands a dynamical explanation. There is no obvious anthropic selection effect that would require $\bar\theta$ to be less than $10^{-3}$ or so.

Lacking an anthropic explanation, any model-building attempt to solve the strong CP problem is incomplete until  at least its new mass scales have been stabilized against radiative corrections. Moreover, if the electroweak hierarchy problem is at least partly solved by physics rather than selection effects, it is interesting to study the solution to strong CP in this larger framework.

Often, it turns out that the various dynamical mechanisms introduced to address these hierarchy problems profoundly alter the physics: radiative stability of scales comes at the price of new constraints on the physics addressing $\bar\theta$. For a well-known example, saxion cosmology places strong constraints on axion models once supersymmetry is introduced to stabilize the scale of Peccei-Quinn breaking and at least partly stabilize the electroweak scale~\cite{Banks:1993en,Banks:1996ea,Banks:2002sd}. Cosmology also constrains composite axion models~\cite{Dobrescu:1996jp}. In Nelson-Barr (NB) models~\cite{nelsoncp,barrcp,barrcp2}, supersymmetry can stabilize the scale of spontaneous CP violation and the electroweak scale, but also introduces new radiative sources for $\bar\theta$~\cite{Dine:1993qm}. Composite NB models are similarly constrained~\cite{Vecchi:2014hpa,dinedraperbn}. 

This is not to say that there cannot be model-building solutions for each of these problems. Viable composite NB models are known~\cite{Vecchi:2014hpa}, for example, and supersymmetric NB is less constrained in gauge-mediated models~\cite{hillerschmaltz,dinedraperbn}. But the mechanisms for stabilizing the scales are generally not innocuous or modular, and incorporating them provides sensitive probe of the robustness of solutions to strong CP.

A number of models have been built attempting to solve the strong CP problem by introducing copies of the Standard Model (SM) related to each other by a \zt~symmetry. These models can be  classified  by the transformation of $SU(3)_c$ under the \zt: 
\begin{enumerate}
\item{$SU(3)_c \rightarrow SU(3)_c$.  In this case, the \zt~is identified with a generalized parity symmetry, which forbids a microscopic $\bar\theta$~\cite{Beg:1978mt,mohapatrasenjanovic,Georgi:1978xz,Babu:1989rb,barrsenjanovic}.}
\item{$SU(3)_c \rightarrow SU(3)_c^\prime$.  In these theories, the generalized parity mechanism is not operative, but $\bar\theta$ may be relaxed at the same time as $\bar\theta'$ by a ``heavy axion"~\cite{Rubakov:1997vp,Berezhiani:2000gh,Fukuda:2015ana} or a ``heavy $\eta^\prime$"~\cite{hook_cp_violation}.}
\end{enumerate}

All such models introduce a new scale associated with the breaking of \zt, and we have argued that at least this scale must be stabilized before the models are complete. What is the impact of the stabilization mechanism on the solutions?

In this paper we study supersymmetric \zt~(``\sz")~models, which can address both the scale of \zt~breaking and the ``big" electroweak hierarchy problem. We focus on generalized parity models, but our basic conclusions about supersymmetry extend in a simple way to heavy axion and heavy $\eta^\prime$ models. In Sections 2.1 and 2.2 we sketch the minimal non-supersymmetric parity models. We demonstrate that corrections to $\bar\theta$ sometimes arise in low orders of perturbation theory, leading to constraints on couplings, while higher-dimension operators lead to constraints on scales. In Sections 2.3 and 2.4 we discuss in more detail the role of naturalness of small numbers and hierarchies in solutions to strong CP and comment on parity conservation from a microscopic point of view.  In Section 3 we add supersymmetry. We show that in general, the addition of SUSY, and in particular the new sources of phases from $\arg H_uH_d$ and its mirrors spoil the parity solution to strong CP.  We also argue that the same problem affects the mirror $\eta^\prime$ model and many heavy axion models. This is not to say that mirror copies of the Standard Model, supersymmetry, and solutions to strong CP cannot coexist, and we briefly consider an interesting example: a supersymmetric twin Higgs model with a Peccei-Quinn (PQ) mechanism to relax $\bar\theta$. Here the \zt~may be present only to solve the little hierarchy problem, and not strong CP. Nonetheless both supersymmetry and the \zt~constrain the implementation of the PQ mechanism: there should be two axion multiplets, one pair for each sector. In Section 4 we summarize and conclude.

%%%%%%%%%%%%%%%%%%%%%%%%%%%%%%%%%%%%%%%%%%%%
%%%%%%%%%%%%%%%%%%%%%%%%%%%%%%%%%%%%%%%%%%%%
%%%%%%%%%%%%%%%%%%%%%%%%%%%%%%%%%%%%%%%%%%%%
\section{Nonsupersymmetric Parity Models}

Refs.~\cite{Beg:1978mt,mohapatrasenjanovic,Georgi:1978xz,Babu:1989rb,barrsenjanovic} considered several classes of models and laid out the basic issues for solving strong CP with spontaneous parity violation. 

In minimal left-right models, $SU(2)_R$ is gauged and transforms under parity as $SU(2)_R\leftrightarrow SU(2)_L$. No additional fermionic matter is added to the SM, and the SM fermions transform in the ordinary way under parity. The Higgs $\Phi$ must transform as a bifundamental under the gauge symmetries, and $\Phi\rightarrow\Phi^\dagger$ under parity. The bare $\theta$ and quark mass matrix $M$ are parity spurions transforming as $\theta\rightarrow-\theta$ and $M\rightarrow M^\dagger$, so $\bar\theta$ vanishes if parity is conserved. 

In these nonsupersymmetric LR models it is not possible to avoid a problematic phase in $\Phi$ without imposing additional structure. Explicit CP violation in the (P-conserving) Higgs potential leads to phase in the gauge invariant combination $\epsilon_{ij}\epsilon_{\alpha \beta} \Phi_{i\alpha}\Phi_{j\beta}$. 

There are two ways to proceed. Mohapatra, Rasin, and Kuchimanchi showed that a supersymmetric left-right model can avoid complex Higgs vevs at tree level~\cite{Mohapatra:1995xd, Kuchimanchi:1995rp}. One-loop phases from the soft SUSY-breaking wino and wino$^\prime$ masses are dangerous because the wino$^\prime$ mass is dominated by the parity-violating vevs, but this can be avoided in UV completions where soft wino masses are automatically real, or more generally if charge conjugation symmetry is imposed in addition to parity~\cite{Mohapatra:1996vg,Kuchimanchi:2010xs}. P and CP-symmetric theories were scrutinized in detail in the supersymmetric context by Mohapatra, Rasin, and Senjanovic~\cite{Mohapatra:1997su}, and it was found that $\bar\theta\approx 0$ could be protected, as long as the scale of the right-handed sector is not more than a few TeV, and C is softly broken at low scale.\footnote{The solution in~\cite{Kuchimanchi:2010xs} with P and CP imposed also works in the non-supersymmetric case.} Such models are therefore attractive from the point of view of experiment, but appear to require new coincidences of scales.

Another solution, not requiring the full imposition of CP, is to add mirror fermions and Higgs bosons and utilize a generalized parity (parity times family symmetry)~\cite{Babu:1989rb,barrsenjanovic}. We will focus on this class of models with mirror matter content and only a parity symmetry. We consider variations in the next sections.

\subsection{Doubling the Electroweak Gauge Group}

Barr, Cheng, and Senjanovic studied a model with complete doubling of the electroweak gauge symmetry, such that $G=SU(3) \times SU(2) \times U(1) \times SU(2)^\prime \times U(1)^\prime$~\cite{barrsenjanovic}.
The particle content is that of the Standard Model under $SU(3) \times SU(2) \times U(1)$, mirrored
under $SU(3) \times SU(2)^\prime \times U(1)^\prime$. For example, a generation of quarks is (in terms of two-component left-handed spinors):
\begin{align}
Q &~(3,2,1,1/6,0),~~\bar u~(\bar 3,1,1,-2/3,0),~~\bar d~(\bar 3,1,1,1/3,0)\;,\nonumber\\
\bar Q^{\prime}&~(\bar 3,1,2,0,-1/6),~~u^{\prime }~(3,1,1,0,2/3),~~d^{\prime }~(3,1,1,0,-1/3)\;.
\end{align}
A \zt ``generalized parity" symmetry interchanges the primed and unprimed gauge groups, changes quarks and leptons into their
mirrors (e.g. $Q\leftrightarrow \bar Q^{\prime *}$), and implements parity on the coordinates.

In addition, there are two Higgs doublets, $\phi$ carrying SM quantum numbers and and $\phi^\prime$ carrying mirror charges. Under \zt, $\phi\leftrightarrow \phi^{\prime*}$. A vev for a singlet pseudoscalar $\sigma$ spontaneously breaks parity somewhere above the electroweak scale. 

The mirror extension of the up-type Yukawa couplings of the theory may be written as
\begin{align}
{\cal L}_{y} = Y^u_{fg} Q_f \phi \bar u_g +  {Y^u_{fg}}^* \bar Q^\prime_f \phi^\prime u^\prime_g + c.c.\;
\label{eq:yukawas}
\end{align}
and similarly for the down-type Yukawas.
A parity-invariant potential is given by
\begin{align}
V=-&m_\phi^2(\vert \phi \vert^2+\vert \phi' \vert^2)+\lambda_\phi(\vert \phi \vert^4+\vert \phi' \vert^4)+\lambda_{\phi\phi^\prime} \vert \phi \vert^2\vert \phi' \vert^2\nonumber\\
-&m_\sigma^2\sigma^2+\lambda_\sigma\sigma^4 + A_{\sigma}\sigma(\vert \phi \vert^2-\vert \phi' \vert^2)+\lambda_{\sigma\phi}\sigma^2(\vert \phi \vert^2+\vert \phi' \vert^2)\;.
\end{align}
The cubic term can split the electroweak scale $|\langle\phi\rangle|=v$ and its mirror scale $|\langle\phi^\prime\rangle|=v^\prime$ such that $v\ll v^\prime\sim\sigma$, but a tree-level tuning is required between $A_{\sigma}\sigma$ and $m^2$. This tuning is another contribution to the electroweak hierarchy problem. The stability of the parity-violating scale $v^\prime$ presents a new possible fine-tuning problem if it is much less than the UV cutoff.

The vevs $v$ and $v^\prime$ may be taken real by separate gauge transformations (equivalently, because hypercharge is doubled, only real gauge-invariant combinations can be constructed from  $\phi$ and $\phi^\prime$.)  Then, at tree level,
\begin{align} 
\arg \det m_q = -\arg \det m_q^\prime
\end{align}
and $\bar\theta$ vanishes, forbidden by generalized parity.

Low order loop corrections are readily seen to preserve this relation.
As noted in~\cite{barrsenjanovic}, the leading corrections to $\bar\theta$ are as in the Standard Model for the primed and unprimed fields separately.  Thus finite contributions only arise at three loop order and are extremely small, while UV divergent terms arise only at seven loops~\cite{ellisgaillard}.  Additional contributions may arise, also at very high order, from the coupling of $\phi$ to $\phi^\prime$.  This is because the unbroken $U(1)$ symmetries forbid mixing between ordinary quarks and their mirrors.  Despite the additional Higgs field, only diagrams with many powers of Yukawa couplings contain phases.

Higher dimension operators, however, can contribute to $\theta$ at tree level, placing an upper bound on the scale of parity violation. At dimension five, the problematic terms are
\begin{align}
{\cal L}\supset {\sigma \over 16 \pi^2 \Lambda_{UV}} F \tilde F+  {\sigma \over \Lambda_{UV}} \left (\gamma_{fg} Q_f \phi \bar u_g  - \gamma_{fg}^* \bar Q^\prime_f \phi^\prime u^{\prime }_g + c.c. \right )
\end{align}
These couplings restrict $\sigma \over \Lambda_{UV}$ to be less than $10^{-10}$; if $\Lambda_{UV}= M_p$, the scale of parity breaking needs to be less than about $10^{8}$
GeV.  Without adding extra structure to the theory,  a massive fine-tuning is required (in addition to the ordinary electroweak hierarchy problem). Indeed, the tuning is worse than the original small-number problem we set out to solve. Even if structure is added so that the leading higher-dimension operators appear at dimension six, the hierarchy problem associated with the scale of parity breaking is equivalent to the tuning needed to solve strong CP by hand.

The model has a virtue over models of spontaneous CP violation, which will be true of the other models we study here as well.  Because CP is violated from the start, no coincidences are required to obtain a large CKM phase. In Nelson-Barr type models, fixing the CKM phase requires potentially disparate numbers (combinations of scales and couplings) be quite close to one another~\cite{Bento:1991ez, Vecchi:2014hpa, dinedraperbn}. However, apart from the issue of tuning,
this model has curious physical features, including a massless dark photon and new massive stable particles (the model possesses an accidental $Z_2$ symmetry under which all of the new fermions are odd). These objects are constrained by astrophysics (stellar and supernova cooling) and cosmology ($N_{eff}$, searches for stable fractionally charged hadrons), and motivate consideration of a variation in which hypercharge transforms trivially under parity~\cite{barrsenjanovic}.

\subsection{Models Without Doubling $U(1)$}

A simple variation on the model of the previous section is to take hypercharge to transform into itself under the \zt. This model was originally discussed in~\cite{barrsenjanovic}, and was reconsidered recently in~\cite{hook_p_violation}, in which it was pointed out that the upper bound on the scale of parity violation from higher dimension operators might place some of the new light states in reach of the LHC.  Removing the second hypercharge avoids some of the troubling phenomenological features of the previous model. However, in the absence of a UV completion it remains highly tuned, and moreover suppressing radiative corrections to $\bar\theta$ poses new challenges.

The renormalizable potential for $\phi$ and $\phi^\prime$ does not admit parity-violating solutions, so at a minimum it is necessary to add a parity odd singlet, $\sigma$ (this is again similar to the case of the previous model), whose expectation value is the origin of parity violation.

As noted in~\cite{barrsenjanovic}, the model allows electroweak singlet mass mixing terms such as $m_u u^\prime \bar u$.  These couplings are CP conserving as a consequence of hermiticity.  The masses 
can naturally be small (in the sense of 't Hooft)
and can play an important role in coupling the new sector to the SM.  In particular, they violate any would-be accidental discrete symmetries of the type found in the previous model.

There is another interesting class of couplings which are allowed, 
\begin{align}
{\cal L}\supset i (y^u)_{fg}\sigma (\bar u_f u_g^\prime - \bar u_g^* u_f^{\prime *})\;,
\label{eq:yuk2}
\end{align}
and similar for the down-type singlet quarks, with hermitian matrices $y^u$ and $y^d$.
These couplings are parity-invariant and CP-violating.

As was argued in~\cite{hook_p_violation}, although the scale of parity violation sets the mass scale for the primed sector, the presence of small Yukawas in~(\ref{eq:yukawas}) suggests that some of the mirror fields may be light enough to see at the LHC. One the other hand, if the couplings of (\ref{eq:yuk2}) are of order one, the new fields will be heavy and the low energy theory is simply the Standard Model. To have states within collider reach, these couplings must also be small (which is again natural in the sense of 't Hooft). 

In fact, radiative stability of $\bar\theta=0$ suggests that the couplings in Eq.~(\ref{eq:yuk2}) must be small. At tree level, $\bar\theta$ still vanishes.  For example, the quark mass matrix has the form:
\beq
{\cal M}_u = \left (\begin{matrix} 0 & {Y_u}^\dagger v^\prime  \cr 
{Y_u} v & m_u+ i \sigma  y_u \end{matrix} \right ).
\eeq
The upper-left block is zero, reflecting the fact that the quarks $u$ and $\bar u^\prime$ carrying $SU(2)$ charge do not mix at leading order. Therefore, the determinant of ${\cal M}_u$ is real at tree level. (This is reminiscent of the Nelson-Barr mechanism, where a zero in the quark mass matrix also plays a crucial role.)  At the one loop level, a non-zero upper left block is generated from diagrams of the form
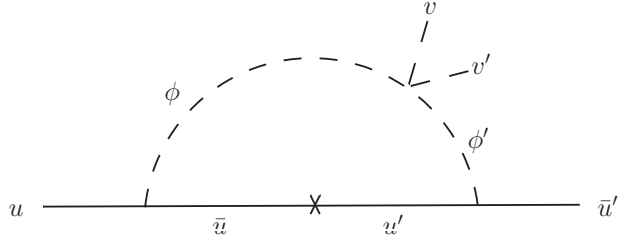
\begin{center}
\fcolorbox{white}{white}{
 \scalebox{.7}{  \begin{picture}(350,135) (86,1)
    \SetWidth{1.0}
    \SetColor{Black}
    \Line(101,9)(390,10)
    \Arc[dash,dashsize=10,clock](245.559,-1.129)(90.13,173.548,7.093)
    \Line[dash,dashsize=10](298,75)(308,109)
    \Line[dash,dashsize=10](299,74)(330,82)
        \Line(250,15)(245,5)
        \Line(245,15)(250,5)
    \Text(83,4)[lb]{\Large{\Black{$u$}}}
    \Text(401,4)[lb]{\Large{\Black{${\bar u}^\prime$}}}
    \Text(307,113)[lb]{\Large{\Black{$v$}}}
    \Text(333,81)[lb]{\Large{\Black{$v^\prime$}}}
    \Text(167,64)[lb]{\Large{\Black{$\phi$}}}
    \Text(331,38)[lb]{\Large{\Black{$\phi^\prime$}}}
    \Text(193,-6)[lb]{\Large{\Black{$\bar u$}}}
    \Text(285,-6)[lb]{\Large{\Black{$u^\prime$}}}
  \end{picture}}
}
\end{center}
but the contribution is proportional to $(m_u+i\sigma y_u)^\dagger$, and does not generate a one-loop $\bar\theta$. There are also one-loop diagrams with a propagating $\sigma$ field that correct the ordinary Yukawa couplings and their mirrors:
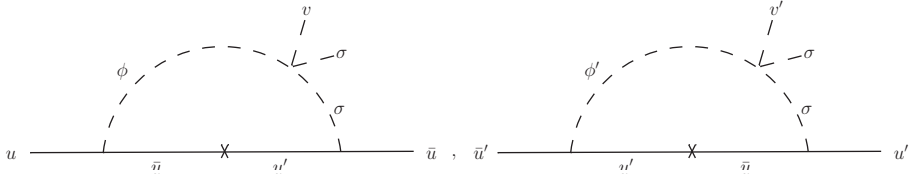
\begin{center}
\fcolorbox{white}{white}{
 \scalebox{.5}{  \begin{picture}(350,135) (86,1)
    \SetWidth{1.0}
    \SetColor{Black}
    \Line(101,9)(390,10)
    \Arc[dash,dashsize=10,clock](245.559,-1.129)(90.13,173.548,7.093)
    \Line[dash,dashsize=10](298,75)(308,109)
    \Line[dash,dashsize=10](299,74)(330,82)
        \Line(250,15)(245,5)
        \Line(245,15)(250,5)
    \Text(83,4)[lb]{\Large{\Black{$u$}}}
    \Text(401,4)[lb]{\Large{\Black{${\bar u}$}}}
    \Text(307,113)[lb]{\Large{\Black{$v$}}}
    \Text(333,81)[lb]{\Large{\Black{$\sigma$}}}
    \Text(167,64)[lb]{\Large{\Black{$\phi$}}}
    \Text(331,38)[lb]{\Large{\Black{$\sigma$}}}
    \Text(193,-6)[lb]{\Large{\Black{$\bar u$}}}
    \Text(285,-6)[lb]{\Large{\Black{$u^\prime$}}}
    \Text(420,4)[lb]{\Large{\Black{,}}}
  \end{picture}
  
   \begin{picture}(350,135) (86,1)
    \SetWidth{1.0}
    \SetColor{Black}
    \Line(101,9)(390,10)
    \Arc[dash,dashsize=10,clock](245.559,-1.129)(90.13,173.548,7.093)
    \Line[dash,dashsize=10](298,75)(308,109)
    \Line[dash,dashsize=10](299,74)(330,82)
        \Line(250,15)(245,5)
        \Line(245,15)(250,5)
    \Text(83,4)[lb]{\Large{\Black{$\bar u^\prime$}}}
    \Text(401,4)[lb]{\Large{\Black{${u^\prime}$}}}
    \Text(307,113)[lb]{\Large{\Black{$v^\prime$}}}
    \Text(333,81)[lb]{\Large{\Black{$\sigma$}}}
    \Text(167,64)[lb]{\Large{\Black{$\phi^\prime$}}}
    \Text(331,38)[lb]{\Large{\Black{$\sigma$}}}
    \Text(193,-6)[lb]{\Large{\Black{$u^\prime$}}}
    \Text(285,-6)[lb]{\Large{\Black{$\bar u$}}}
  \end{picture}
  
  }
}
\end{center}
These diagrams provide a correction to $\bar \theta$ of order
\begin{align}
\Delta_{\bar\theta}\sim\frac{\lambda_{\phi\sigma} y}{16\pi^2 }\frac{m}{m_{parity}}\;,
\end{align}
where $m$ and $y$ are characteristic elements of $m_u$ and $y_u$ and we have approximated the scale of the diagrams as $m_{parity}\sim v^{\prime }\sim\sigma\gg v$.
The scalar potential couplings are not protected by any symmetry,  and if they are $\mathcal{O}(1)$, then we require 
$y\times m/\sigma \lesssim 10^{-8}$. Of course, this may be viewed as simply a natural requirement on models of this type, since discrete symmetries are restored in the limit $y,m \rightarrow 0$.

%%%%%%%%%%%%%%%%%%%%%%%%%%

At two loops, there are also contributions to $\bar\theta$ from diagrams such as:
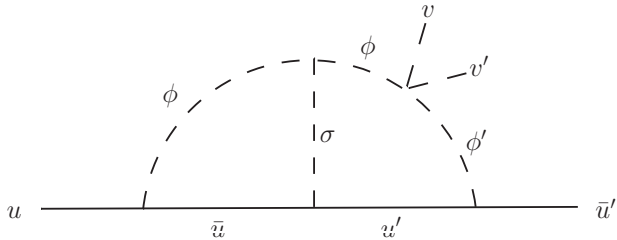
\begin{center}
\fcolorbox{white}{white}{
 \scalebox{.7}{  \begin{picture}(350,135) (86,1)
    \SetWidth{1.0}
    \SetColor{Black}
    \Line(101,9)(390,10)
    \Arc[dash,dashsize=10,clock](245.559,-1.129)(90.13,173.548,7.093)
    \Line[dash,dashsize=10](248,90)(248,10)
    \Line[dash,dashsize=10](298,75)(308,109)
    \Line[dash,dashsize=10](299,74)(330,82)
    \Text(83,4)[lb]{\Large{\Black{$u$}}}
    \Text(401,4)[lb]{\Large{\Black{${\bar u}^\prime$}}}
    \Text(307,113)[lb]{\Large{\Black{$v$}}}
    \Text(333,81)[lb]{\Large{\Black{$v^\prime$}}}
    \Text(252,45)[lb]{\Large{\Black{$\sigma$}}}
    \Text(167,64)[lb]{\Large{\Black{$\phi$}}}
    \Text(274,90)[lb]{\Large{\Black{$\phi$}}}
    \Text(331,38)[lb]{\Large{\Black{$\phi^\prime$}}}
    \Text(193,-6)[lb]{\Large{\Black{$\bar u$}}}
    \Text(285,-6)[lb]{\Large{\Black{$u^\prime$}}}
  \end{picture}}
}
\end{center}
There is no analogous correction to the lower-right block of the mass matrix, so this diagram gives a contribution to $\bar\theta$ of order
\begin{align}
\Delta_{\bar\theta}\sim\frac{y^2  \lambda_{\phi\phi^\prime}}{(16\pi^2)^2}\frac{A_\sigma}{m_{parity}}\;.
\end{align}
%The second power of $y$ reflects that the diagonal block of ${\cal M}_u$ that is nonzero at tree level is of order $y$. 
If the scalar potential couplings are $\mathcal{O}(1)$, then we require 
$y \lesssim 10^{-3}$. Again this may be viewed as a natural requirement on these models.

\subsection{Naturalness and Technical Naturalness}
Perhaps it is worth belaboring the various roles of naturalness in solutions to strong CP. Small $\bar\theta$ is technically natural in the Standard Model~\cite{ellisgaillard}.\footnote{We distinguish technical naturalness of a small parameter, where the radiative corrections to the parameter are not larger than its magnitude, and naturalness in the sense of 't Hooft, where a symmetry is restored when the parameter is taken to zero. Small parameters that are natural in the sense of 't Hooft are technically natural, but not vice versa.} More precisely, suppose that the effective theory below an ultraviolet scale $\Lambda_{SM}$ consists only of SM degrees of freedom. Then the correction to $\bar\theta$ from modes below $\Lambda_{SM}$ is additive: $\bar\theta=\bar\theta_0+\Delta^{SM}_{\bar\theta}$, where $\bar\theta_0$ is the boundary condition at $\Lambda_{SM}$. The correction $\Delta^{SM}_{\bar\theta}$ is tiny and independent of $\bar\theta_0$. If $\bar\theta_0\simeq 0$, $\bar\theta$ remains small at low energies.

Strong CP is then a problem of the ultraviolet completion, including in particular all of the physics of baryogenesis, but also of flavor, dark matter, neutrino masses, inflation, quantum gravity, and any mechanisms addressing the electroweak hierarchy problem. Why should all of these ingredients preserve ${\bar\theta_0}\simeq 0$ without fine-tuning?

The Nelson-Barr mechanism, and the \zt-based mechanisms like generalized parity, introduce new physics at a scale $\Lambda_{\bar\theta}$ designed to provide symmetry-based explanations for the boundary condition $\bar\theta_0=0$.  $\Lambda_{SM}\approx \Lambda_{\bar\theta}$ is presumed to lie somewhere above the electroweak scale, but small compared to UV scales like the Planck or unification scales. These mechanisms generally operate at tree level with the inclusion only of marginal operators. On the other hand, in the parity model discussed in the previous section, and also in minimal NB models~\cite{Bento:1991ez}, the quantum correction to the low energy $\bar\theta$, $\Delta_{\bar\theta}$, includes threshold corrections at $\Lambda_{\bar\theta}$ that are potentially large. In particular, among the new fields responsible for P or CP violation, typically some couplings $y_{ij}$  must be small in order to prevent unacceptably large $\Delta_{\bar\theta}$. The smallness of the $y_{ij}$ may be technically natural, but it is not immediately clear that much has been gained relative to the SM: a small-number problem of order $10^{-10}$ in one boundary condition has been traded for several small numbers of order $10^{-4}$ or less in other boundary conditions. However, the smallness of the $y_{ij}$ may also be natural in the more restrictive sense of 't Hooft, possibly representing an improvement over the technical naturalness of small $\bar\theta_0$ in the SM.

To properly compare these models to the SM, we must include the other ``new physics" ingredients listed above. There are three issues. First, does $y_{ij}\ll 1$ remain natural? If small $y_{ij}$ is natural in the sense of 't Hooft, the relevant symmetry must simply be extended to the UV completion. 

The second issue is whether the other physical ingredients give new contributions to $\Delta_{\bar\theta}$, apart from those associated with the couplings $y_{ij}$. Typically this is a question of scales. If $\Lambda_{\bar\theta}$ is low compared to the other scales $\Lambda_i$ of new physics, the theory at $\Lambda_{\bar\theta}$ may be well-approximated by, e.g., the generalized parity model above plus some higher-dimension operators suppressed by the $\Lambda_{i}$. If the operators are of high dimension, or if the ratios $\Lambda_{\bar\theta}/\Lambda_i$ are sufficiently small, $\Delta_{\bar\theta}$ is safe, and again we have improved on the situation in the SM. However, among the different modules of new physics, mechanisms addressing the electroweak hierarchy play a special role: if the electroweak scale is completely natural, then there is new physics below $\Lambda_{\bar\theta}$. In that case we can draw no conclusions without evaluating $\Delta_{\bar\theta}$ on a model-by-model basis, including both the dynamics setting $\bar\theta=0$ at $\Lambda_{\bar\theta}$, and the dynamics protecting the electroweak scale below $\Lambda_{\bar\theta}$.

The third issue is the naturalness of the scale $\Lambda_{\bar\theta}$ itself. We have argued that this scale must be stabilized, so again a proper assessment of $\Delta_{\bar\theta}$ and the naturalness of small couplings must incorporate additional dynamics. Indeed, keeping $\Delta_{\bar\theta}$ small will in general introduce extra requirements on new couplings. We will discuss the constraints on a supersymmetric stabilization of $\Lambda_{\bar\theta}$ in detail in Section 3.

\subsection{Comment on Microscopic Considerations}
\label{sec:UV}
Apart from issues of tuning and radiative stability in the models discussed above, we can ask what might be required in order for a microscopic theory to conserve a parity symmetry.

 In the context of Nelson-Barr models, Ref.~\cite{dinedraperbn} challenged the notion that CP conservation at the level of the underlying laws necessarily implies a vanishing ``bare," or high scale, $\theta$.  There are two issues illustrated by properties of string theory.
 
First, string theories typically possess a CP symmetry, but also have moduli, some of which are odd under CP.  To conserve CP,  the odd moduli (many of which can be thought of as axions) must be stabilized at the origin. The assumption of the spontaneous CP violation scenarios, in this context, is that the axions are much heavier than the conventional QCD axion of Peccei and Quinn.  Thus the magnitude of the ``bare" $\theta$ becomes a problem of dynamics rather than symmetry.  
 
Second, in the context of a flux landscape, it is necessary that all fluxes odd under CP vanish.  This typically occurs only in an exponentially small fraction of states.
 
 The same statements apply to parity in string theories.  In string compactifications, one can specify conditions under which parity is conserved (in the four dimensional sense).  Again, there are typically moduli, of which of order a half are parity odd, and all of which must vanish (or be stabilized at scales far below $M_p$).  Similarly, of order a half of possible fluxes will be parity odd.  So, from a string point of view, if fluxes are important in determining the structure of the vacuum, it is not at all obvious that parity conservation in the underlying equations leads to vanishing $\theta$ at very high scales.

%%%%%%%%%%%%%%%%%%%%%%%%%%%%%%%%%%%%%%%%%%%%
%%%%%%%%%%%%%%%%%%%%%%%%%%%%%%%%%%%%%%%%%%%%
%%%%%%%%%%%%%%%%%%%%%%%%%%%%%%%%%%%%%%%%%%%%
\section{Supersymmetry}
\subsection{Generalized Parity Models}

We have enumerated three challenges to building models that solve the strong CP problem through spontaneous parity violation:
\begin{enumerate}
\item  Higher-dimension operators force a low scale of P violation, implying a fine tuning of new fundamental
scalar masses in addition to the ordinary electroweak scale.
\item For generic values of the couplings, radiative corrections can generate a substantial $\bar\theta$ near the parity-violating scale in some models.
\item  The assumption that P accounts for a small bare $\bar\theta$ is sensitive to ultraviolet physics, such as moduli stabilization in string theory.
\end{enumerate}
These issues are also present in Nelson-Barr models that solve strong CP with a particular form of spontaneous CP violation, and it is instructive to recall the impact of supersymmetry~\cite{Dine:1993qm,hillerschmaltz,dinedraperbn}. If SUSY is broken below the CP scale, then it solves (at least) the new hierarchy problems. Furthermore holomorphy of the superpotential typically reduces or eliminates dimension-five operators, allowing a higher scale of P violation, and nonrenormalization protects $\bar\theta$ until the SUSY breaking scale. The third issue is not addressed with SUSY (but without committing to a complete ultraviolet theory, the severity of the problem is also not clear.)

On the other hand, supersymmetry introduces new potential problems in NB models. At leading order, SUSY introduces new sources of $\bar\theta$. For example, in the Minimal Supersymmetric Standard Model,
\begin{align}
\bar\theta=\theta-\arg\det (y_u y_d) -3 \arg (m_{\tilde g}) -3 \arg (v_u v_d)\;,
\end{align}
so, barring cancellations, we require the gluino mass and $B_\mu$ to be real to a part in $10^{10}$. This suggests that the CP-violating sector must be sequestered from the SUSY-breaking sector in the UV, so that the SUSY-breaking $F$-term is real. Even then, the anomaly-mediated contribution to the gluino mass, $m_{\tilde g}\sim W_0^*/M_p^2$, can be problematic.\footnote{The phase of $W_0$ is also related to the UV questions discussed in Sec.~\ref{sec:UV}.} There are also new radiative corrections to $\arg\det(y_uy_d)$ at the soft SUSY-breaking scale, particularly in the presence of $A$-terms and squark flavor violation, which generically obtain complex values after integrating out the CP-violating sector~\cite{Dine:1993qm}.  The problems of radiative corrections from $A$-terms and flavor violation, as well as the anomaly-mediated gluino mass, are substantially ameliorated in low-scale SUSY-breaking with real $F$ and gauge mediation.

Supersymmetric models of spontaneous P violation exhibit similar features as well as new challenges.  
To illustrate them, we adopt a simple model containing a mirror $SU(2)$ and mirrors of all of the chiral superfields of the MSSM, including mirror Higgs doublets $H_u^\prime$ and $H_d^\prime$. 
On the Higgs multiplets, \zt~acts as $H_{u,d}\rightarrow H_{u,d}^{\prime*}$. The vacuum angle in this model is
\begin{align}
\bar\theta=\theta-\arg\det (y_u y_d)-\arg\det (y_u^\prime y_d^\prime) -3 \arg (m_{\tilde g}) -3 \arg (v_u v_d)-3 \arg (v_u^\prime v_d^\prime)\;.
\end{align}
To spontaneously break parity, we can, for example, add a singlet $\chi$ with superpotential couplings:\footnote{This superpotential, and the \zt-symmetric soft parameters listed below, can also provide a partial model for the supersymmetric twin Higgs mechanism~\cite{Falkowski:2006qq}. For the purpose of solving the little hierarchy problem it maybe useful to also double QCD and explicitly break the \zt, in which case the parity solution to strong CP is not available. We will comment briefly on strong CP in twin Higgs models below.}
\begin{align}
W_\chi=  \chi (\lambda H_u H_d + \lambda^* H_u^\prime H_d^\prime - M^2)\;.
\end{align}
At the classical level, there is a large degeneracy of vacua.  We focus on the vacuum in which the gauge-invariant combination
$H_u^\prime H_d^\prime$ has a (complex) expectation value:
\beq
\lambda^* H_u^\prime H_d^\prime = M^2\;,~~\chi = 0.
\eeq
This vacuum breaks parity. 

Beyond scale stabilization, the most immediate advantage of supersymmetry is in the problem of higher-dimension operators. The leading terms now appear at dimension six, for example,
\begin{align}
\int d^2\theta \lambda \frac{(H_uH_d+ H_u^\prime H_d^\prime)}{M_p^2}W_\alpha^2+c.c.
\label{eq:W}
\end{align}
The scale of parity violation might therefore be as high as $10^{13}$ GeV, rather than $10^8$ GeV in non-SUSY models.

However, new problems arise with the breaking of supersymmetry.  Below the scale $\sqrt{F}$ of spontaneous SUSY-breaking, we can introduce soft masses with scale $m_{soft}$, and as in the supersymmetric Nelson-Barr models, $m_{soft}$ should be less than the scale of parity violation $M$ to solve the new hierarchy problems in that sector. $\sqrt{F}$ can still be larger than $M$, in which case the soft terms are parity-symmetric. For example, soft scalar mass terms
\begin{align}
m_{H_u}^2 (\vert H_u \vert^2 +\vert H_u^\prime \vert^2)+m_{H_d}^2( \vert H_d \vert^2 + \vert H_d^\prime \vert^2)
\label{eq:softm}
\end{align}
can stabilize $H_{u,d}$, while only leading to small shifts in the $H_{u,d}^\prime$ expectation values for $m_{soft}\ll M$. There are also soft breaking $B_\mu$ terms of the form
\begin{align}
B_\mu H_u H_d + B_\mu^* H_u^\prime H_d^\prime + {\rm c.c.}
\end{align}
Again the impact on the $H_{u,d}^\prime$ vevs is small. On the other hand, these terms play a critical role in determining $H_{u,d}$. For general $\beta\equiv \arg(B_\mu)$,  $\alpha\equiv\arg(H_u H_d)$ differs from  $\alpha^\prime\equiv\arg(H_u^{\prime*} H_d^{\prime*})$. Minimizing the low-energy potential, to leading order in $m_{soft}/M$, they can be shown to be related by
\begin{align}
\tan(\alpha-\alpha^\prime)=\left(\frac{m_{H_u}^2+m_{H_d}^2}{4|B_\mu|}\right)\sin(\alpha^\prime+\beta)
\end{align}
As a result, for general soft parameters,
\begin{align}
\bar\theta\sim1\;.
\end{align}

We stress that the problem of the $B_\mu$ phase is general in supersymmetric parity models attempting to solve strong CP. Unlike their nonsupersymmetric counterparts, supersymmetric parity models possess complex gauge invariant order parameters $H_uH_d$ and $H_u^\prime H_d^\prime$. The former is fixed by soft SUSY breaking, while the latter is fixed by parity breaking, and the phases are uncorrelated in general.

Radiative corrections are also potentially dangerous when they are sensitive to parity violation. Because the wino$^\prime$ receives most of its mass from SUSY-preserving parity violation, phases in the soft wino and wino$^\prime$ masses generate $\arg\det m_q$ at one loop (this problem affects the minimal left-right models~\cite{Mohapatra:1995xd,Mohapatra:1996vg}). If there are complex $A$ terms, there are similar one-loop contributions to $\arg(m_{\tilde g})$ from quark-squark loops with one $A$-insertion.  

We can ask whether phases in gaugino masses, $B_\mu$, and other soft breaking parameters might realistically be small.  To study this question we need a model for supersymmetry breaking and its mediation.   
Supersymmetry breaking need not introduce additional phases, at least at tree level.  For example, in an O'Raifeartaigh sector,
\beq
W_{or} = \lambda Z (A^2 - \mu^2) + m AB\;,
\eeq
parity invariance ($Z,A,B \rightarrow Z^*,A^*,B^*$) implies real parameters. 
However, mediation typically introduces many phases.  Again, it is important to recall that if a generalized parity symmetry solves strong CP, all phases allowed by parity should be included, unless new structure in addition to parity is added to suppress them.  In gravity mediation, essentially all of the soft breakings are associated with such complex couplings.  One exception is the gluino mass: $\int d^2 \theta a Z W_\alpha^2$
is real for the O'Raifeartaigh type model above.  But couplings like:
\begin{align}
\int d^4 \theta & Z^\dagger ( b H_u H_d + b^* H_u^\prime H_d^\prime)+ \int d^4 \theta Z^\dagger Z (c H_u H_d + c^* H_u^\prime H_d^\prime)\nonumber\\
&+\int d^4 \theta Z^\dagger Z (d_{f g} Q_f Q^\dagger_g + d_{fg}^* Q_f^\prime Q_g^{\prime\dagger}) + \dots\;.
\label{eq:bmu}
\end{align}
contribute phases to the $\mu$ and $B_\mu$ terms, the $A$ terms, and the Hermitian squark mass matrices. As we have discussed, all of these phases are potentially dangerous for $\theta$, either at tree level or one loop.  
Maintaining $\bar\theta=0$ thus presents an even more severe problem in parity models than in the case of supersymmetric Nelson-Barr, where CP alone was enough to suppress most phases in the absence of phases in $Z$.

In NB models, gauge mediation appears much more promising for protecting $\bar\theta$~\cite{hillerschmaltz,dinedraperbn}. If there are no phases in the hidden sector, the leading contributions to the Hermitian squark and slepton masses are real, the wino and wino$^\prime$ masses are real, and the $A$ terms are suppressed. This argument carries through in supersymmetric parity models, in particular eliminating or suppressing many of the dangerous one-loop corrections.
However, gauge mediation is incomplete until an additional mechanism is added to generate $\mu$ and $B_\mu$, which necessarily requires couplings beyond the gauge interactions. These couplings always have a chiral structure, as for example in Eq.~(\ref{eq:bmu}), and thus in general introduce new phases. The problem of $ H_u H_d$ and  $H_u^\prime H_d^\prime$ phase misalignment remains.

To summarize, supersymmetry provides an obvious stabilization of scales in generalized parity models, but offers only limited improvement over the other problems encountered in the non-supersymmetric case, and raises new, severe difficulties with maintaining $\bar\theta=0$.

%%%%%%%%%%%%%%%%%%%%%%%%%%%%%%%%%%%%%%%%%%%%
\subsection{Heavy Axion and $\eta^\prime$ Models}
Two other examples of non-supersymmetric \zt~models -- in which $SU(3)_c$ is also copied -- are heavy axion models~\cite{Rubakov:1997vp,Berezhiani:2000gh,Fukuda:2015ana} and the heavy $\eta^\prime$ model of~\cite{hook_cp_violation}. In both cases, the \zt~requires the UV $\bar\theta$ angles in the two sectors to match. In the heavy axion case, an axion is introduced with \zt-preserving couplings to the topological term,
\begin{align}
\mathcal{L}\supset\left(\frac{a}{f}-\bar\theta\right)\times\left(G\tilde G+G'\tilde G'\right)\;.
\label{eq:heavyaxion}
\end{align}
Now the axion potential comes predominantly from the characteristic scale of the primed sector $\Lambda'$, which differs from $\Lambda$ once \zt~is broken at a high scale. The axion can be much heavier in these models, circumventing astrophysical constraints on $f_a$, and in particular its mass may not be suppressed by any light quark mass if there are none in the mirror sector. However, it still dynamically cancels $\bar\theta$ as long as there are no large radiative corrections that spoil the relation $\bar\theta=\bar\theta'$ below the scale of \zt~breaking. In minimal non-supersymmetric models, this may indeed be the case.

Heavy $\eta^\prime$ models behave similarly. With a high scale of \zt-breaking, all mirror quarks are heavy, and $SU(3)_c^\prime$ runs strong at $\Lambda^\prime\gg\Lambda$. An extra pair of massless quarks ($\psi,\bar\psi$), bifundamental under $SU(3)_c\times SU(3)_c^\prime$, condense at $\Lambda^\prime$ and spontaneously break chiral symmetries. Chiral $SU(3)$ breaking leads to a light set of pseudogoldstones in the adjoint of ordinary QCD. Breaking of the anomalous chiral $U(1)$ leads to a scalar ``$\eta^{\prime\prime}$," analogous to the ordinary $\eta^\prime$. Since the chiral $U(1)$ is anomalous under $SU(3)_c^\prime$, the $\eta^{\prime\prime}$ obtains a mass of order $\Lambda^{\prime}$. Since the chiral $U(1)$ is also anomalous under $SU(3)_c$, the $\eta^{\prime\prime}$ couples to the topological term of ordinary QCD. Schematically, the $\eta^{\prime\prime}$ is controlled by a potential of the form
\begin{align}
\mathcal{L}\supset\left(\frac{\eta^{\prime\prime}}{f_{\pi^\prime}}-\bar\theta\right)\times G\tilde G-m_{\eta^{\prime\prime}}^2\left(\frac{\eta^{\prime\prime}}{f_{\pi^\prime}}-\bar\theta\right)^2+\cdots\;.
\end{align}
The quadratic term fixes $\eta^{\prime\prime}$ at $f_{\pi^\prime}\bar\theta$, simultaneously canceling the QCD vacuum angle.

Both of these models invoke a new scale associated with spontaneous \zt~breaking, and thus both require a stabilization mechanism. One possibility is to promote them to \sz~models. However, it is easy to see that adding supersymmetry typically contaminates these models with the same problems as in the generalized parity case. In particular, a large phase difference between $H_uH_d$ and $H_u'H_d'$ is expected at tree level, for exactly the same reason: $H_uH_d$ and $H_u^\prime H_d^\prime$ are controlled by different physics. Likewise, unless the SUSY-breaking sector is accidentally CP-conserving and SUSY-breaking is mediated by something like gauge mediation, large one-loop contributions to the visible sector's $\arg\det(m_q)$ and $\arg(m_{\tilde g})$ arise from diagrams with insertions of the complex wino and wino$^\prime$ soft masses and $A$-terms.

Unlike the parity case, the problem in these models is not that $\bar\theta\neq 0$, but rather that $\bar\theta\neq \bar\theta^\prime$. In the approximation $\Lambda^\prime\gg\Lambda$, the dynamics of QCD$^\prime$ fixes the axion/$\eta^{\prime\prime}$ such that $\bar\theta^\prime$ is cancelled, e.g. $a=f\bar\theta^\prime$. After axion/$\eta^{\prime\prime}$ stabilization, the effective $\bar\theta$ becomes, for example,
\begin{align}
\bar\theta\rightarrow \bar\theta-a/f = \bar\theta-\bar\theta^\prime\;.
\end{align}

Our discussion of heavy axions includes, in particular, \zt-symmetric versions of the string axion, which can in principle avoid the well-known problem of PQ-breaking higher-dimension operators. However, there is at least one interesting axion model that provides a partial exception. The \zt-symmetric Weinberg-Wilczek axion model of~\cite{Berezhiani:2000gh} can be straightforwardly supersymmetrized. In this model, one linear combination of $\arg(H_uH_d)$ and $\arg(H_u^\prime H_d^\prime)$ is fixed at zero, while the other plays the role of the axion and couples in a \zt-symmetric way as in~(\ref{eq:heavyaxion}). For this reason the model is elegantly exempt from the problem of a large tree-level phase difference between $H_uH_d$ and $H_u^\prime H_d^\prime$. (Another way to say it is the relevant PQ symmetry forbids $B_\mu$.) In this model, the scale of spontaneous \zt- and PQ-breaking is determined by SUSY-breaking terms in the potential, and a variety of experimental constraints require the former scales to be in the range $10^4-10^5$ GeV.\footnote{The constraints are assessed in the nonsupersymmetric case in~\cite{Berezhiani:2000gh}, but as they concern only very light degrees of freedom, they are applicable more generally.} Therefore a complete solution to the electroweak hierarchy problem is not obtained, but it is at least substantially reduced. On the other hand, since \zt-breaking gives an order-one splitting between the masses of squarks and mirror squarks, winos and mirror winos, etc., there are still one-loop corrections to $\bar\theta-\bar\theta^\prime$ unless SUSY-breaking is gauge mediated and $F$ is real.

%%%%%%%%%%%%%%%%%%%%%%%%%%%%%%%%%%%%%%%%%%%%
%%%%%%%%%%%%%%%%%%%%%%%%%%%%%%%%%%%%%%%%%%%%
%%%%%%%%%%%%%%%%%%%%%%%%%%%%%%%%%%%%%%%%%%%%
\subsection{Strong CP and Supersymmetric Twin Higgs}

Twin Higgs models also introduce mirror-type symmetries, including copies of the electroweak group, but for the different purpose of addressing the little hierarchy problem. What is the status of strong CP in the twin Higgs context? 

Let us briefly recall how the twin mechanism operates~\cite{Chacko:2005pe}. The \zt~symmetry automatically yields an accidental $SU(4)$ symmetry in the Higgs sector mass terms, under which $(H,H^\prime)$ transforms as a fundamental. Spontaneous breaking of $SU(4)$ at a scale $f$ produces pseudo-Goldstone bosons.
The \zt~forces radiative corrections to the mass terms to respect $SU(4)$, so there are no quadratically divergent corrections to the pseudo-Goldstone masses. 
%However, since the $SU(4)$ is broken explicitly by Yukawa and gauge interactions, generically there are also $SU(4)$-breaking quartic couplings in the potential. 
The SM Higgs doublet should be mostly aligned with Goldstone directions, and the observed Higgs boson should have SM-like properties. Therefore, $f$ should be a \zt-breaking scale, pointing primarily in the direction of the mirror $SU(2)$. This can be achieved with explicit soft \zt-breaking (and/or spontaneous breaking of \zt~at $f$, as in the supersymmetric model of Eq.~(\ref{eq:W})). 
The electroweak scale $v$ can be naturally somewhat smaller than $f$, by a ratio of couplings, while $f$ can be naturally smaller than the UV cutoff $\Lambda$ by $\sim 4\pi$. The little hierarchy problem is relaxed when $\Lambda\sim 5$ TeV.\footnote{For a particularly clear recent presentation of the tuning properties in twin Higgs models, see Sec. III of~\cite{Craig:2015pha}.} Full protection of the scale $f$ requires a UV completion, such as supersymmetry~\cite{Falkowski:2006qq,Chang:2006ra}.\footnote{Supersymmetric twin Higgs models with \zt-symmetric soft masses have the additional feature that the electroweak scale is ``doubly protected." As usual, quadratic divergences cancel because of supersymmetry. Furthermore, terms proportional to $m_{soft}^2\times\log(\Lambda/m_{soft})$ know about SUSY breaking, but not about \zt-breaking, and therefore cannot give mass to the Goldstone~\cite{Falkowski:2006qq,Chang:2006ra}.}

As we have noted, the twin Higgs \zt~symmetry is suggestive of the mechanisms discussed in previous sections for addressing strong CP. Indeed, the superpotential and soft masses we have studied in Eqs.~(\ref{eq:W},\ref{eq:softm}) have precisely the SU(4)-symmetric structure of the SUSY twin higgs model in~\cite{Falkowski:2006qq}. 

In fact, the demands on the \zt~for the purposes of addressing a little hierarchy are much weaker than the demands on \zt~symmetries when used in solutions to strong CP. The light quarks and leptons (not to mention $\theta$) are irrelevant to the little hierarchy, and in the minimal twin model the first and second generation mirror fields would simply be omitted, corresponding to a hard breaking of the \zt~\cite{Craig:2015pha}.  Furthermore, electroweak fine-tuning grows with $f$ in twin models, so models with full \zt~copies predict new light states. Cosmologically it is convenient to omit them~\cite{Chacko:2005pe,Barbieri:2005ri,Chang:2006ra}. Explicit \zt~breaking can also be introduced to raise the light Higgs mass in SUSY twin Higgs~\cite{Craig:2013fga}. With various sources of \zt~breaking, including different quark spectra in the two sectors, there would be little reason to expect cancellation between, for example, $\arg\det m_q$ and $\arg\det m_q^\prime$ in generalized parity models.

Therefore, we can distinguish two cases. In the first, a \zt~symmetry of very high quality plays a role in addressing both the little hierarchy problem and strong CP. It is attractive to imagine that the mirror world might serve dual purposes. This scenario suggests a nonstandard cosmology, as well as a non-supersymmetric UV completion, since we have seen that it is difficult to maintain small $\bar\theta$ in \sz~models.  In the second scenario, a \zt~symmetry (either exact or approximate) plays a role in the electroweak hierarchy problem, but is unrelated to strong CP. For the latter, the Peccei-Quinn mechanism~\cite{Peccei:1977hh,Peccei:1977ur} remains an option.

Let us briefly consider the features of axions in supersymmetric twin Higgs models. It has been argued that the twin Higgs \zt~should be extended to map $SU(3)_c\rightarrow SU(3)_c^\prime$ in order to cancel large two-loop contributions to the Higgs mass parameters~\cite{Chacko:2005pe,Craig:2015pha}. In the absence of light mirror quarks, or in the presence of large contributions to $\bar\theta-\bar\theta^\prime$ in supersymmetric models, a single axion is disfavored, since its potential
\begin{align}
V(a)\sim m \Lambda^3 \cos(a/f_a-\bar\theta)+m^\prime \Lambda^{^\prime 3} \cos(a/f_a-\bar\theta^\prime)
\end{align}
relaxes neither of the two angles when they are different. (This is equivalent to a ``heavy axion" solution of the type discussed above, although with $f$ not much greater than a few times $v$, the scales of the two contributions to the axion potential are not widely split.) Likewise the heavy $\eta^\prime$ solution is ill-suited to this setup, predicting both an order-1 effective vacuum angle and an inconveniently light scalar octet under QCD.

Therefore, the most plausible setting for the Peccei-Quinn mechanism in SUSY twin Higgs models is the case where \zt~acts nontrivially on the axion multiplet, with $a\rightarrow a^\prime$. The potential becomes
\begin{align}
V(a)\sim m \Lambda^3 \cos(a/f_a-\bar\theta)+m^\prime \Lambda^{^\prime 3} \cos(a^\prime/f_a-\bar\theta^\prime)
\end{align}
and both vacuum angles are relaxed by their own axions.

As in the non-twin setting, the primary issues with this mechanism are the quality of the PQ symmetry and the saxion problems~\cite{Banks:1993en,Banks:1996ea,Banks:2002sd}. This issues are in tension with one another: the former is suggestive of a string axion, which typically is weakly coupled (large $f_a$), while the latter is suggestive of a more strongly coupled axion. One possibility is that the axion is a string axion, but the scale of SUSY-breaking is high, $m_{3/2}\sim 100$ TeV, so that the cosmological issues with the saxion are avoided. In a twin Higgs setting two saxions would accompany the two axions. The primary observation is that the minimal twin mechanism does not fully protect the little hierarchy if $m_{3/2}$ is so high; however, it may still improve on the tuning with supersymmetry alone.

%
%
%
%%%%%%%%%%%%%%%%%%%%%%%%%%%%%%%%%%%%%%%%%%%%
%%%%%%%%%%%%%%%%%%%%%%%%%%%%%%%%%%%%%%%%%%%%
%%%%%%%%%%%%%%%%%%%%%%%%%%%%%%%%%%%%%%%%%%%%
\section{Conclusions}

In this paper we have explored some of the theoretical issues surrounding solutions to strong CP that implement \zt~copies of the Standard Model.  The smallness of $\bar\theta$ demands a dynamical solution that does not introduce more small number problems than it solves, so a crucial element to any solution is the radiative stability of new scales. In \zt~models, the scale of \zt-breaking must be far below the Planck scale to suppress contributions to $\bar\theta$ from higher-dimension operators. Although supersymmetry can stabilize the scale of \zt~breaking, we have argued that \sz~models are extremely constrained by new contributions to $\bar\theta$ at tree level and one loop after \zt~and SUSY are broken. Our observations affect models with generalized parity symmetries, heavy axions, and heavy mirror $\eta^\prime$s.

We have confined our analysis to cases where supersymmetry stabilizes the hierarchy and have said nothing about the possibility of a strong-dynamics origin for the \zt~scale. It would be interesting to assess this mechanism further (along the lines of~\cite{Vecchi:2014hpa} in the context of the Nelson-Barr mechanism). We have also commented briefly on strong CP in models where a \zt~symmetry plays a role in stabilizing the electroweak scale.
The status of \zt~solutions to strong CP in nonsupersymmetric UV completions of twin Higgs models, as well as the phenomenology of two axions (and saxions) in supersymmetric twin Higgs, is deserving of further study.

\vspace{1cm}

%{\bf  Acknowledgements:} 
\noindent {\bf Acknowledgements:}  This work was supported in part by the U.S. Department of Energy grant number DE-FG02-04ER41286. PD thanks Nathaniel Craig for explaining the twin Higgs mechanism to him, and Simon Knapen for helpful discussions on strong CP in twin Higgs and for comments on the manuscript.  A.H. wishes to thank Santa Cruz for Particle Physics for its warm hospitality and  Adam Coogan and Tim Stefaniak for conversations. AA's travel was sponsored by University of Hail.

\bibliography{p_violation_refs}{}
\bibliographystyle{jhep}

\end{document}